# Investigation on the Coronal Magnetic Field Using a Type II Solar Radio Burst

V. Vasanth[1*] · S. Umapathy[1,3] · Bojan Vršnak[2] · Tomislav Žic[2] · O. Prakash[1]

[1]School of Physics, Madurai Kamaraj University, Madurai – 625021, India

[2]Hvar Observatory, Faculty of Geodesy, Zagreb, Croatia

[3]Radio Astronomy Center, NCRA-TIFR, Ooty, India

*email: vasanth_velu2007@yahoo.co.in

## Abstract

The Type-II solar radio burst recorded on 13 June 2010 by the radio spectrograph of the Hiraiso Solar Observatory was employed to estimate the magnetic-field strength in the solar corona. The burst was characterized by a well pronounced band-splitting, which we used to estimate the density jump at the shock and Alfvén Mach number using the Rankine–Hugoniot relations. The plasma frequency of the Type-II bursts is converted into height [$R$] in solar radii using the appropriate density model, then we estimated the shock speed [$V_s$], coronal Alfvén velocity [$V_a$], and the magnetic-field strength at different heights. The relative bandwidth of the band-split is found to be in the range 0.2 – 0.25, corresponding to the density jump of $X = 1.44 – 1.56$, and the Alfvén Mach number of $M_A = 1.35 – 1.45$. The inferred mean shock speed was on the order of $V \approx 667$ km s$^{-1}$. From the dependencies $V(R)$ and $M_A(R)$ we found that Alfvén speed slightly decreases at $R \approx 1.3 – 1.5$. The magnetic-field strength decreases from a value between 2.7 and 1.7 G at $R \approx 1.3 – 1.5$ R$_\odot$ depending on the coronal-density model employed. We find that our results are in good agreement with the empirical scaling by Dulk and McLean (Solar Phys. **57**, 279, 1978) and Gopalswamy *et al.* (Astrophys. J. **744**, 72, 2012). Our result shows that Type-II band splitting method is an important tool for inferring the coronal magnetic field, especially when independent measurements were made from white light observations.

## 1. Introduction

Payne-Scott *et al.* (1947) identified the Type-II radio bursts in the meter-wavelength range and later Wild and McCready (1950) named the slow drifting radio bursts Type-II solar radio bursts to differentiate them from the fast drifting Type-III bursts. Type-II radio bursts show a slow drift in time from high to low frequency with a drift rate typically ≤ 0.5 MHz s$^{-1}$ at metric wavelengths (Mann, Klassen, and Classen, 1996) with decreasing values at longer wavelengths (Gopalswamy, 2006). The radio emission is caused by the conversion of plasma waves excited by electrons accelerated at MHD shocks propagating through the solar corona (Nelson and Melrose, 1985). In the radio dynamic spectrum, Type-II bursts usually appear as two emission bands, which correspond to the local plasma frequency and its harmonic. Type-II bursts have been used as an important tracer of coronal and interplanetary shock waves.

Type II bursts also show band splitting in both the fundamental and harmonic bands. One of the interpretations is that band splitting is a result of the plasma emission from the upstream and downstream shock regions (Smerd, Sheridan, and Stewart, 1974; Vršnak *et al*. 2001 and references therein). The frequency difference between the two split bands is thus related to the shock-compression ratio, which provides an estimate of the Alfvén Mach number [$M_A$] according to the Rankine–Hugoniot relations. If the ambient density and shock speed are known, the ambient Alfvén speed and hence the magnetic-field strength can be determined (Vršnak *et al*. 2002; Cho *et al*. 2007). In this respect it should be noted that multilane band emission can be also a consequence of the emission being excited at two or more locations along the shock front. Yet, in "classical" band-split Type-II bursts, the two emission lanes show similar morphology, similar intensity variations, and correlated frequency drift, implying that the emission comes from a common radio-source trajectory (for details, see Vršnak *et al*. 2001). Furthermore, Vršnak *et al*. (2001) have shown that in the case of interplanetary Type-II bursts, the extrapolation of the band-split emission maps to the density jump at the *in-situ* recorded shock (see Figure 4 of Vršnak *et al*. 2001). Both aspects strongly support the interpretation in terms of emission coming from the upstream and downstream regions of the shock.

The magnetic field plays an important role in the solar corona; it is considered to be the main factor for coronal heating, particle acceleration, and formation of structures such as prominences and coronal mass ejections. The coronal magnetic-field strength has been determined in the past by several methods. At photospheric levels, the conventional Zeeman splitting of spectral lines in the visible part of spectrum is employed to determine the strength of the magnetic field, while at coronal levels the magnetic field is reconstructed by extrapolating the photospheric field, most often applying the potential-field approximation. Estimates of the magnetic-field strength using microwave, decimeter and meter-wave radio burst had been performed by applying indirect methods that involve interpretation of data in terms of the gyro-frequency. Recently, Gopalswamy and Yashiro (2011) proposed a novel method of estimating the coronal magnetic field from the standoff distance of the CME driven shock, measured in white-light coronagraphic images. The magnetic-field strength inferred by employing Type-II radio bursts at meter wavelengths has been determined by several authors (see, e.g., Smerd, Sheridan, and Stewart, 1974; Karlicky and Tlamicha, 1974, Vršnak *et al*. 2002; Subramanian, Ebenezer, and Raveesha, 2010; Cho *et al*. 2007; Ma *et al*. 2011; Gopalswamy *et al*. 2012).

In this study we apply the band-split technique to the metric Type-II burst recorded on 13 June 2010. This event has been extensively studied because the CME-driven shock was directly detected at EUV wavelengths (Patsourakos, Vourlidas, and Stenborg, 2010; Kozarev *et al*. 2011; Ma *et al*. 2011; Gopalswamy *et al*. 2012). In particular Gopalswamy *et al*. (2012) used this event to derive the coronal magnetic field by combining EUV observations from the *Solar Dynamics Observatory (SDO)* and band-splitting information of the associated metric Type-II burst. Observing such events is rare because of the limited field of view of the SDO EUV instrument. We exploit this rare opportunity to evaluate the use of coronal-density models by comparing the results obtained from density models and direct measurements.

## 2. Event Description

The radio dynamic spectrum recorded on 13 June 2010 by the Hiraiso radio spectrograph is shown in Figure 1 (see also Gopalswamy *et al*. 2012). The burst was also recorded by the Culgoora Radio Observatory. The burst showed both harmonic and fundamental emission bands, the latter being much weaker than the former. The harmonic started at 320 MHz at 05:37:10 UT and ended at ≈ 60 MHz at 05:46 UT. Overall characteristics of the Type-II burst are summarized in Table 1.

The associated GOES M1.0 X-ray flare, located at S25W84, started at 05:35 UT, seven minutes before the onset of the Type-II burst. The soft X-ray flux peaked two minutes after the Type-II burst onset. Also, a coronal mass ejection (CME) was reported in the SOHO/LASCO CME catalog (cdaw.gsfc.nasa.gov: Gopalswamy *et al*. 2009), first observed in the LASCO field of view at 06:06 UT at $R$ = 2.57 R$_\odot$, where $R$ is the radial distance expressed in solar radii. It was a slow CME, having a mean plane-of-the-sky speed of 320 km s$^{-1}$ and weakly accelerating at the rate of +2.58 m s$^{-2}$ in the LASCO field of view. The Type-II emission appears only in the metric domain and no longer-wavelength Type-II was found in the *Wind/WAVES* dynamic spectrum.

The event on 13 June 2010 was observed by the *Atmospheric Imaging Assembly (AIA) onboard SDO*, which takes full-disk images in ten wavelengths with arcsecond spatial resolution and 12-sec and cadence. Patsourakos, Vourlidas, and Stenborg, (2010) studied this event as an expanding CME bubble and determined its kinematics. Kozarev *et al*. (2011) reported the implications of the shock for the acceleration of solar energetic particles (SEPs). Ma *et al*. (2011) reported the magnetic-field strength from the band-splitting technique and used the Sittler and Guhathakurta (1999) electron-density model to find the Type-II speed by assuming the quasi-perpendicular shock condition. Gopalswamy *et al*. (2012) combined the EUV shock measurements (standoff distance, the radius of curvature of the CME, and the shock speed) and combined them with the band-splitting measurement of the Type-II burst to derive the magnetic field. We applied the electron-density models (Newkirk and Saito electron-density models) to find the shock speed and to derive the Alfvén speed and magnetic field. This provides an opportunity to test the utility of the electron-density models when direct measurement of the shock speed is not available.

In the earlier reports (Kozarev *et al*. 2011; Gopalswamy *et al*. 2012) it is pointed out that at 5:37 UT, a shock wave appears, which travels faster and becomes get separated from the CME in the AIA FOV, while the Type-II burst was observed by the radio observatories at 5:37 UT, which coincides with the observation. The wave reached the AIA FOV edge at 5:42 UT, was followed by a CME at 5:44 UT, and left the AIA FOV. From Figure 5 of Gopalswamy *et al*. (2012) the Alfvén speed first increases and then decreases; and the decreasing of the Alfvén speed indicates the decrease in shock speed. The shock is weakening at this time and seems to dissipate when the CME reaches the LASCO FOV. In our analysis Figure 2 shows the evolution of X and $M_A$, from Figure 3 and 4 the shock speed and Alfvén speed decreases and reaches a minimum, which indicates the shock is very weak and not seen at the time the CME reached the LASCO FOV. Therefore no longer- wavelength Type-II burst was observed.

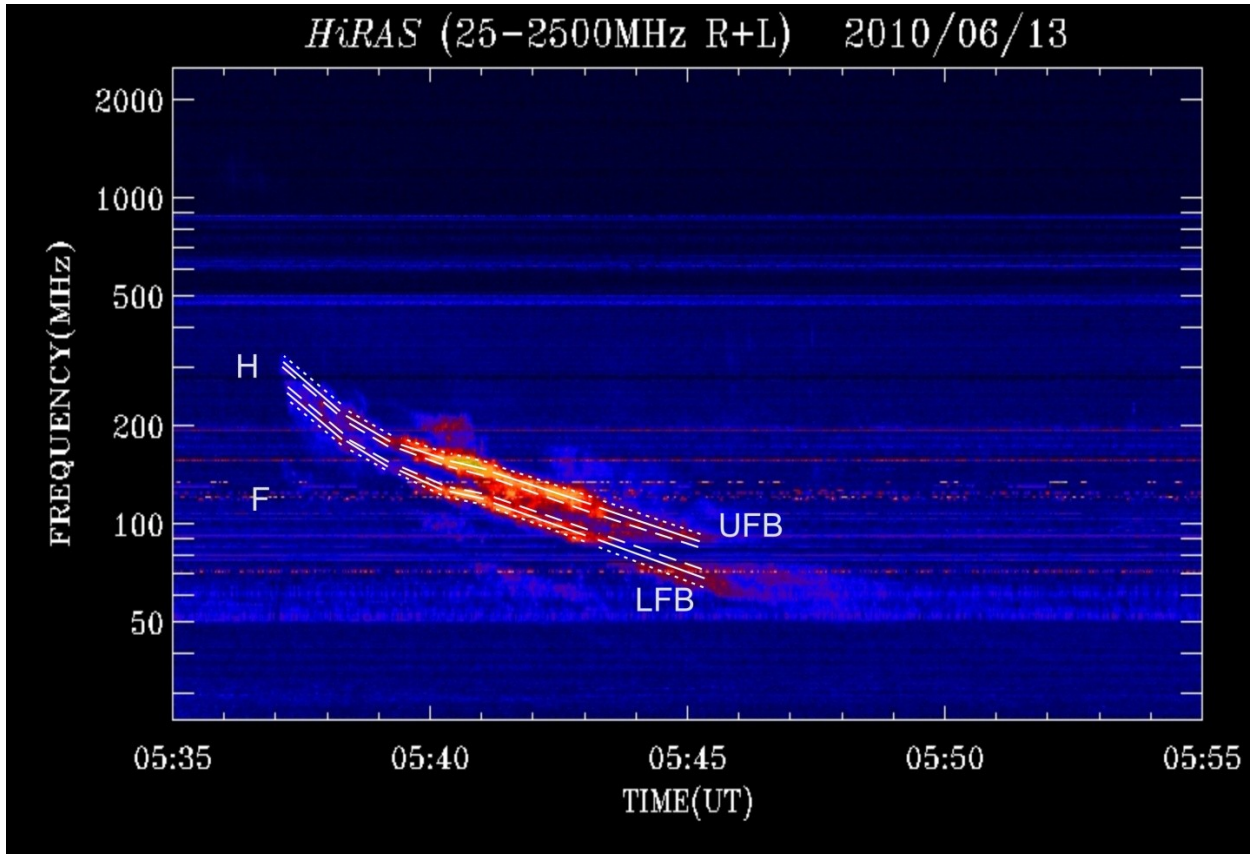

**Figure 1.** Spectrum of the Type-II burst recorded on 13 June 2010 by the Hiraiso Solar Observatory (H – harmonic band; F – fundamental band; UFB – upper frequency branch; LFB – lower frequency branch). Solid lines follow maximum intensities in the emission lanes, whereas dotted and dashed lines represent an estimate of the outer and inner borders of emission, respectively. [The band-splitting is clear between 5:39 – 5:43 UT]

## 3. Method

**Table 1: Overall characteristics of the type II burst**

| | |
|---|---|
| Start time – End time [*UT*] | 05:37 – 05:46 |
| frequency range | 320 – 60 MHz |
| Mean drift rate, $\Delta f / \Delta t$ [*H-band*] | 0.18 MHz s$^{-1}$ |
| Mean relative band splitting [$\Delta f / f$] | 0.23 |
| Mean density jump [*X*] | 1.52 |

Using the relation between the plasma frequency [$f$] and electron density [$n$] given by $f = 9 \times 10^3 n^{1/2}$ MHz, the Type-II bursts observed at a local plasma frequency [$f$] can be used to get information about the plasma density [$n$] at which the emission takes place. The region ahead of the shock (upstream region) is characterized by the electron density [$n_1$] and plasma frequency [$f_1$]. The emission from this region creates the lower frequency branch (LFB) of the split band. The region behind the shock (downstream region) is compressed, so the electron density [$n_2$] is higher than [$n_1$] and the corresponding plasma frequency [$f_2$] is higher than [$f_1$]. The emission from the downstream region corresponds to the upper frequency branch (UFB) of band split. The relative instantaneous bandwidth (BDW) of the splitting can be expressed as:

$$\text{BDW} = \Delta f / f = (f_2 - f_1)/f_1 = (n_2/n_1)^{1/2} - 1 , \qquad (1)$$

Thus, the density jump [$X$], across the shock can be written as

$$X = n_2/n_1 = (\text{BDW} + 1)^2 . \qquad (2)$$

From the density jump [$X$] we derive the Alfvén Mach number [$M_A$] using a simplified Rankine–Hugoniot jump relation:

$$M_A = (X(X+5) / 2(4-X))^{1/2}, \qquad (3)$$

which is valid for perpendicular shocks in a low plasma-to-magnetic pressure ratio environment ($\beta \ll 1$) and the specific-heat ratio (the polytropic index) of $\gamma = 5/3$ (for a discussion see Vršnak et al. 2002).

We used the emission frequency and the drift rate of the harmonic band to estimate the height of the radio source and the shock speed, employing various coronal electron density models. Once the shock speed [$V_s$] is estimated, it is possible to convert the Alfvén Mach number to the Alfvén speed using the relation

$$V_a = V_s / M_A . \qquad (4)$$

We determine the ambient magnetic-field strength [$B$] using the Alfvén speed [$V_a$] and the LFB frequency [$f_1$]:

$$B = 5.1 \times 10^{-5} \times f_1 V_a , \qquad (5)$$

where the frequency is expressed in MHz, Alfvén velocity in km s$^{-1}$, and the magnetic field in gauss.

Finally, we check the consistency of shock compression ratio [$n_2/n_1$] obtained from band splitting to the theoretical value. From Figure 1 we obtain the compression ratio and compare them with the theoretical value in Table 2. Assuming that the shock is quasi-perpendicular at low heights, we can use the formula by Drane and Mckee, (1993):

$$n_2/n_1 = 2(\gamma + 1)/\{D + [D^2 + 4(\gamma + 1)(2 - \gamma) M_A^{-2}]^{1/2}\}, \qquad (6)$$

Where $D \approx (\gamma-1) + (2/M_s^2 + \gamma/M_A^2)$ and $M_s = V_s/C_s$, $M_s$ – sonic Mach number, $M_A$ – Alfvénic Mach number and $\gamma$ – adiabatic index. For a 2MK corona, the sound speed is $C_s \approx 128$ km s$^{-1}$ (Gopalswamy *et al.* 2012), so $M_s \approx 6$ for the derived shock speed. The shock compression values obtained using Equation (6) are consistent with the values obtained from the radio spectrum, and with plasma $\beta = C_s^2/V_a^2$ using the values of $C_s \approx 128$ km s$^{-1}$ and $V_a$ in Table 2.

If the shock is parallel, then the density compression ratio can be found using (Drane and McKee, 1993)

$$n_2/n_1 = (\gamma + 1)/\{(\gamma - 1) + (2/M_A^{-2})\} \qquad (7)$$

From this we can derive the Mach number as $M_A^2 = n_2/n_1$.

The values in Table 2 measured from the Type-II burst observed by the Hairiso radio spectrograph are compared with direct measurements from SDO/AIA (Gopalswamy *et al.* 2012), whose plasma frequency [$f_p$] is 78, 64, 59, and 52 MHz at time 5:39 – 5:42 UT. We find that the observed frequencies are the same (78, 59 and 52 MHz at 5:39, 5:41 and 5:42 UT) but the radial distance obtained from the electron-density model, the corresponding shock speed, the Alfvén speed and the magnetic field slightly vary from the results of Gopalswamy *et al.* (2012). Table 2 lists the measured properties of Type-II burst for perpendicular and parallel shock condition; Table 2 compares the Mach number, Alfvén speed, and estimated magnetic-field strength for parallel and perpendicular shock for the times during the Type-II bursts band split. At low height in the corona the emission follow a quasi-perpendicular condition (which are consistent with earlier results) at most location, rather than a parallel condition, which would give slightly higher values. Kozarev *et al.* (2011) reported that the magnetic field geometry above the active region that produced the CME on 13 June 2010 was closed, So that the shock might have been quasi-perpendicular at the CME nose that produced the Type-II radio burst.

## 4. Analysis and Results

We measured the band splitting of the harmonic band, and employed Equations (1) – (3), to estimate the relative band splitting [BDW], density jump [X] and Alfvén Mach-number [$M_A$] as a function of time. The procedure of the band-split measurements is illustrated in Figure 1, where the full lines follow maximum intensities in the UFB and LFB emission lanes of the harmonic band, whereas the dotted and dashed lines represent an estimate of the outer and inner borders of this emission, respectively. In the following, we use the measurements along the full lines; the dotted and dashed lines will be used for an estimate of the uncertainity. The Type-II burst interval is divided into 30-second sub-intervals defined by times $t_{i=1}$ to $t_N$. At each of these N instants the emission frequencies of LFB and UFB is measured, providing evaluation of the instantaneous bandwidth as a function of time, [BDW(t)].

**Table 2:** Evolution of the type II burst for perpendicular and parallel shock condition

| Time [UT] | $f_p$ [MHz] | $N_e$ $10^7$ [cm$^{-3}$] | $R/R_\odot$ | $V_S$ [km/s] | Perpendicular shock condition | | | | | | | Parallel shock condition | | | | |
|---|---|---|---|---|---|---|---|---|---|---|---|---|---|---|---|---|
| | | | | | $M_A$ | $V_a$ [km/s] | $M_s$ | B [G] | $N_2/N_1$ (Radio) | $N_2/N_1$ (Theory) | β | $M_A$ | $V_a$ [km/s] | B [G] | $N_2/N_1$ (Theory) | β |
| 5:38 | 99 | 12.0 | 1.24 | | | | | | | | | | | | | |
| 5:39 | 78 | 7.5 | 1.32 | 818 | 1.35 | 606 | 6.38 | 2.41 | 1.44 | 1.41 | 0.04 | 1.2 | 682 | 2.71 | 3.72 | 0.04 |
| 5:40 | 66 | 5.3 | 1.39 | 603 | 1.42 | 425 | 4.71 | 1.43 | 1.53 | 1.46 | 0.09 | 1.24 | 486 | 1.63 | 3.52 | 0.06 |
| 5:41 | 59 | 4.3 | 1.43 | 580 | 1.43 | 406 | 4.53 | 1.22 | 1.54 | 1.47 | 0.09 | 1.24 | 468 | 1.40 | 3.49 | 0.07 |
| 5:42 | 52 | 3.3 | 1.49 | | | | | | | | | | | | | |

In Table 2, column 1 denotes the time of observation of the radio burst in UT, column 2 represents the plasma frequency [$f_p$] in MHz, column 3 specifies the electron density [$N_e$], column 4 defines the radial distance $R/R_\odot$ of corresponding to the plasma frequency [$f_p$], column 5 indicates the estimated shock speed estimated by applying electron density model (1×

Newkirk model) and columns 5 - 12 denotes the measured properties of perpendicular shock condition, specifically column 6 denotes the Mach number [$M_A$], Column 7 represents the Alfvén speed [$V_A$], column 8 denotes the sonic Mach number [$M_s$], column 9 represents the estimated magnetic field [B (G)], column 10 and 11 compares the shock compression ratio [$n_2/n_1$] from the radio and theory and Column 12 denotes the plasma β values. Similarly the measured properties of Type-II bursts for parallel shock condition are given in columns 13 – 17. Column 13 denotes the Mach number [$M_A$], Column 14 represents the Alfvén speed [$V_A$], column 15 represents the estimated magnetic field [B (G)] column 16 compares the shock compression ratio [$n_2/n_1$] from the theory with the radio listed in column 10 and Column 17 denotes the plasma β values.

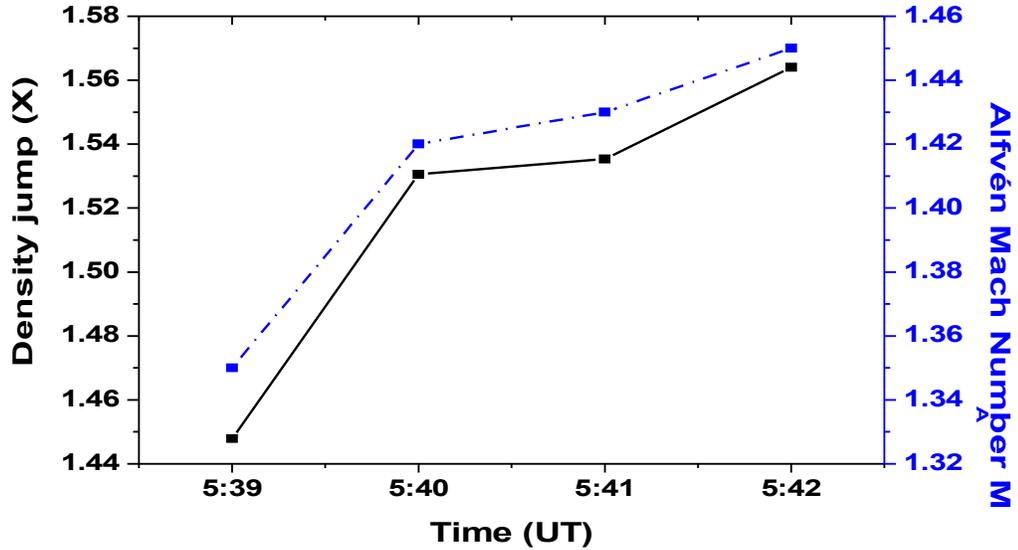

**Figure 2.** Density jump [*X*] (full line) and Alfvén Mach-number [$M_A$] (dashed line) estimated from the band splitting of the harmonic band.

The evolution of the UFB and LFB emission, measured by following the intensity maximum depicted by full lines in Figure 1, as well as the corresponding values of *X* and $M_A$, are shown in Table 2. The evolution of *X* and $M_A$ is also presented graphically in Figure 2. BDW is in the range 0.2 – 0.25, with the mean value of 0.23, which is within the range found in earlier studies (e.g., Mann, Classen, and Aurass, 1995; Mann, Klassen, and Classen, 1996; Vršnak *et al*. 2002). These values correspond to *X* = 1.44 – 1.56, with a mean value *X* ≈ 1.52, and $M_A$ = 1.35 – 1.45, with a mean value $M_A$ ≈ 1.41.

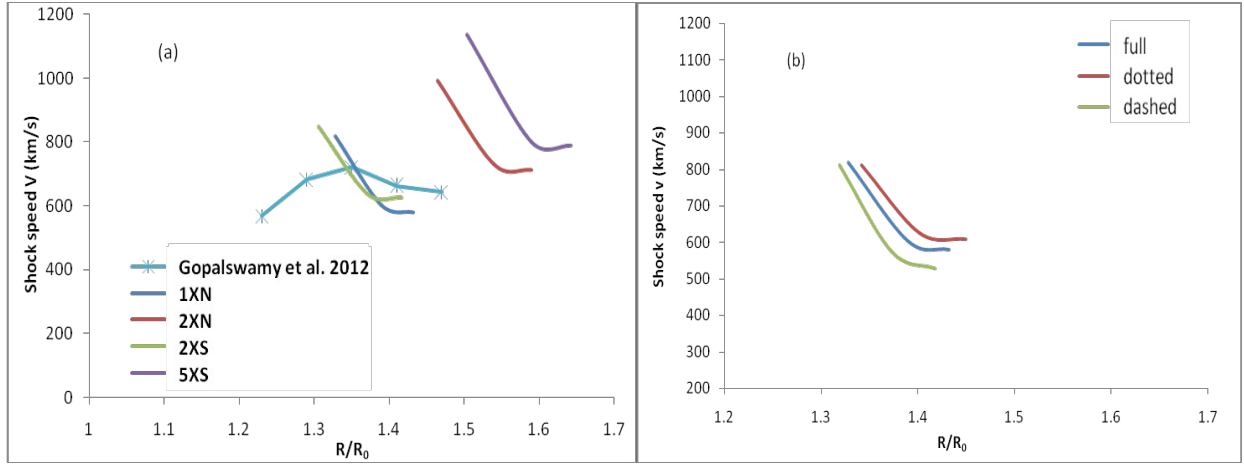

**Figure 3.** The shock speed [$V_s$] as a function of the heliocentric distance [$R$ expressed in units of the solar radius: $R_\odot$]. a) Shock speeds estimated using two- and five-fold Saito (1970) electron-density model and one- and two-fold Newkirk (1961) models are distinguished and the shock speed of Gopalswamy *et al.* (2012) from AIA observation is compared with our result. b) Shock speeds estimated by measuring the full, dashed, and dotted lines in Figure 1, and applying the one-fold Newkirk model.

The coronal electron density is a function of height [$n_e(R)$] decreasing with height, when a suitable electron-density model is applied. We can derive the height of the emission frequency. This is then converted to shock speed, employing electron-density models by Saito (1970) and Newkirk (1961), the two models most appropriate to analyze dynamic spectra of radio bursts associated with eruptive phenomena in active regions, since the Newkirk model is valid in the inner corona within the height R = 1– 3 $R_\odot$, while the Saito model is valid between R = 1– 10 $R_\odot$ and can be compared with the Newkirk electron-density model. The evolution of the shock speed with heliocentric distance (expressed in solar radii) is drawn in Figure 3. In Figure 3a we present the results obtained using two- and five-fold Saito density model as well as one- and two-fold Newkirk model and our result is compared with that of Gopalswamy *et al.* (2012); we find that the shock speed slightly decreases from 818 – 580 km s$^{-1}$ which is consistent with Ma *et al.* (2011) who find the speed of the Type-II shock using the density model proposed by Sittler and Guhathakurta (1999). In Figure 3b we show the results obtained by measuring full, dashed, and dotted lines in Figure 1, and applying the one-fold Newkirk model. From the graphs we find that the shock speed slightly decreases with distance. The mean speed of the radio source is estimated to be 704 and 909 km s$^{-1}$ for two- and five-fold Saito models, respectively. Similar values, on the order of 667 and 812 km s$^{-1}$, are obtained for the one- and two-fold Newkirk model (roughly corresponding to two- and five-fold Saito model). The estimated speed by one-fold Newkirk model and two-fold Saito model are closest to the measured speed (varied by 1.1 – 1.2 times) reported by Gopalswamy *et al*. (2012), while the speeds by two-fold Newkirk and five-fold Saito model varies by a factor of 1.4 – 1.6 times the reported values of Gopalswamy *et al*. (2012).

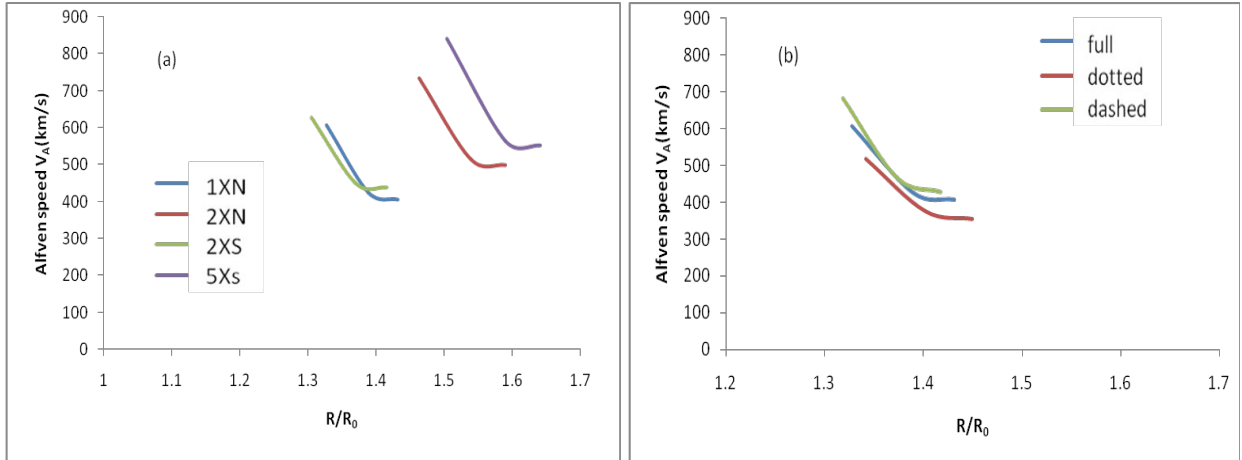

**Figure 4.** Alfvén speed [$V_a$] derived from the Type-II frequency drift and band split, presented as a function of heliocentric distance. a) Values estimated using two- and five-fold Saito (1970) electron-density model and one- and two-fold Newkirk (1961) model. b) Values estimated by measuring the full, dashed, and dotted lines in Figure 1, and applying the one-fold Newkirk model.

Figure 4 shows the Alfvén speed estimated using the inferred shock speed and Mach number. Applying the two- and five-fold Saito, and the one- and two-fold Newkirk density model, we find that inferred values of the Alfvén speed range between 350 – 1000 km s$^{-1}$ in the radial distance range $R$ = 1.3 – 1.7 R$_\odot$, depending not only on the density model employed, but also on the heliocentric distance Figure 4(a). Figure 4(b) shows that the effect of different density models is considerably larger than the effects of different measurement procedures and accuracy of measurements. Obviously, the estimate of Alfvén speed depends strongly on the applied electron density model: higher density implies a higher value of $V_a$.

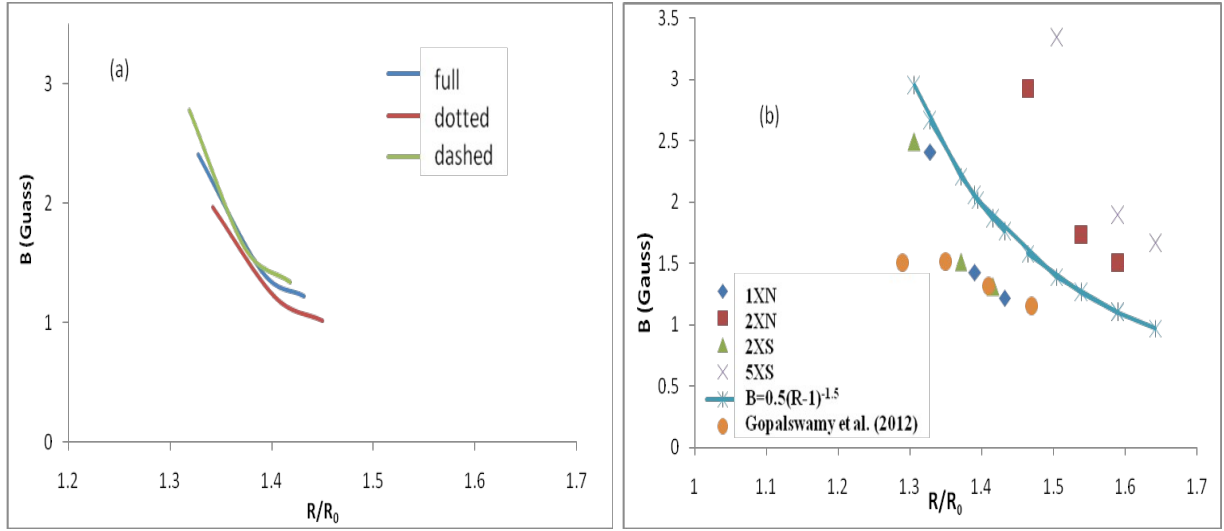

**Figure 5.** Coronal magnetic-field strength presented as a function of heliocentric distance. a) Values estimated by measuring the full, dashed, and dotted lines in Figure 1, and applying the one-fold Newkirk model. b) Comparison of the results for the two- and five-fold Saito and one- and two-fold Newkirk model, with functions $B = 0.5\,(R-1)^{-1.5}$ and the magnetic-field B (G) of Gopalswamy et al. (2012) is compared with our result.

Figure 5 shows the magnetic-field strength estimated using the Alfvén speed and the density in the upstream region determined by LFB frequencies (see Section 3). Details of this method and validation analysis are presented by Vršnak *et al*. (2001) and Cho *et al*. (2007). The magnetic-field strength varies between 3.5 and 1 G in the distance range $R = 1.3 – 1.7\ R_\odot$, depending on the density model employed, as well as on the heliocentric distance (Figure 5a). Figure 5(a) shows that the effect of different density models is considerably larger than the effects of different measurement procedures and accuracy of measurements. However, inspecting Figures 5(a) and (b) one finds that the magnetic field decreases with the distance, regardless of the density model applied or the details of the measurement procedure. It decreases from a value between 3.5 and 1 G at $R \approx 1.3 – 1.7\ R_\odot$ depending on the coronal density model employed.

Vršnak *et al*. (2001) used the Newkirk, Saito, and Leblanc electron-density models to find the shock speed of m-/DH-type-II bursts and Cho *et al*. (2007) used the Newkirk model to find the shock speed of metric (m-) Type-II bursts.

In Figure 5 the dependence $B(R)$ obtained by applying the two-fold Saito and one-fold Newkirk density model is compared with scaling $B = 0.5\,(R-1)^{-1.5}$ proposed by Dulk and McLean (1978), which is appropriate for the coronal magnetic field above active regions. We find that our results for two-fold Saito and one-fold Newkirk model gives values lower than Dulk and

McLean (1978), while five-fold Saito and two-fold Newkirk electron density model gives values which are highly similar to those from Dulk and McLean.

When we compared our results with Gopalswamy *et al*. (2012), we find that the two-fold Saito and one-fold Newkirk model gives values varied by a less than a factor of two, while five-fold Saito and two-fold Newkirk electron density model gives values varied by more than a factor of two. Therefore the results from two-fold Saito and one-fold Newkirk model are found be consistent with the measurements of Gopalswamy *et al*. (2012) who derived the magnetic-field strength without employing the density model. Also Bemporad and Mancuso (2010) showed that the Newkirk model agrees well with the Saito model multiplied by a factor of 2.5, which is also consistent with our result shown in Figure 5.

## 5. Conclusion

We applied the band-splitting method to determine the Alfvén speed and magnetic field in the radial distance range $R = 1.3 - 1.7$ $R_\odot$. Our results are in agreement with other estimates in this height range only for certain density models. Over this height range, we found the relative bandwidth of the band split in the range $\Delta f / f = BDW \approx 0.2 - 0.25$, which is consistent with earlier results: e.g. Vršnak *et al*. (2001) found $BDW = 0.05 - 0.6$ and Mann, Classen and Aurass (1995) and Mann, Klassen, and Classen (1996) found $BDW = 0.1 - 0.7$. The density jump lies between $X = 1.44 - 1.56$ which is in agreement with the earlier reports (Vršnak *et al*. 2002; $X \approx 1 - 2$ and Cho *et al*. 2007 $X = 1.5 - 1.6$). The Alfvén Mach number obtained for this event lies in the range $M_A = 1.35 - 1.45$ and is also in agreement with the earlier reports (Smerd, Sheridan, and Stewart (1974); Vršnak *et al*. 2002; Cho *et al*. 2007; Magdalenic *et al*. 2008, 2010).

From Figure 5, we find that the magnetic-field strength estimated using the two-fold Saito and one-fold Newkirk model gives values that varied by less than a factor of two, while five-fold Saito and two-fold Newkirk electron density model gives values varied by more than a factor of two. Therefore the result from two-fold Saito and one-fold Newkirk model are found be consistent with the measurements of Gopalswamy *et al*. (2012) who derived the magnetic field strength without employing the density model. The magnetic field strength decreases monotonously from a value between 3.5 and 1 G at $R \approx 1.3 - 1.7$ $R_\odot$. Such a result is in good agreement with the scaling proposed by Dulk and McLean (1978) and Gopalswamy *et al*. (2012). At low coronal heights the emission seems to be consistent with quasi-perpendicular conditions rather than the parallel condition. We conclude that the Type-II burst band-split method is an important diagnostic tool for inferring the coronal Alfvén speed and magnetic-field characteristics, especially when the evolution of the shock can be followed using independent measurements using white-light observations (Gopalswamy and Yashiro, 2011) or EUV observations (Gopalswamy *et al*. 2012). The frequency drifting of a Type-II radio burst reflects its driver height under the assumption of a coronal density model. The direct measurement of the driver height from the geometrical properties of the associated CME does not require the use of a coronal density model (Gopalswamy *et al*. 2012). Using the radio data alone requires the use of coronal-density models, which can lead to large uncertainties, as we demonstrated in this work.

## Acknowledgement

We are grateful to the referee whose constructive comments led to significant improvement of the paper. We thank N. Gopalswamy for his suggestions and help to improve the manuscript. The authors are thankful to the Hiraiso Solar Observatory (Hiraiso radio spectrograph) teams for providing the catalog of Type-II bursts. We express our thanks to the online data centers of NOAA and NASA for providing the data. VV thanks UGC for a BSR fellowship in science for meritorious students.

## References

Bemporad, A., Mancuso, S.: 2010, Astrohys. J. **720**:130. doi:10.1088/0004-637X/720/1/130.

Cho, K.-S., Lee, J., Gary, D.E., Moon, Y.-J., Park, Y.D.: 2007, Astrophys. J. **665**, 799. doi:10.1086/519160.

Dulk, G.A., McLean, D.J.: 1978, Solar Phys. **57**, 279. doi: 10.1007/BF00160102.

Drane, B.T., Mckee, C.F.: 1993, Annu. Rev. Astron. Astrophys. **31**, 373. doi:10.1146/annurev.aa.31.090193.002105.

Gopalswamy, 2006, In: Gopalswamy, N., Mewaldt, R., Torsti, J. (eds.) Solar Eruptions and Energetic Particles, AGU Geophys. Monogr. **165**, 207.doi:10.1029/165GM20.

Gopalswamy, N., Yashiro, S., Michalek, G., Stenborg, G., Vourlidas, A., Freeland, S., Howard, R.A.: 2009, Earth Moon Planets, 104, 295. doi: 10.1007/s11038-008-9282-7.

Gopalswamy, N., Yashiro, S.: 2011, Astrophys. J. Lett. **736**, L17. doi:10.1088/2041-8205/736/1/L17.

Gopalswamy, N., Nitta, N., Akiyama, S., Makela, P., Yashiro, S.: 2012, Astrophys. J. **744**, 72. doi:10.1088/0004-637X/744/1/72.

Karlicky, M., Tlamicha, A.: 1974, Bull. Astron. Inst. Czech. **30**, 246.

Kozarev, K.A., Korreck, K.E., Lobzin, V.V., Weber, M.A., Schwadron, N.A.: 2011, Astrophys. J. Lett. **733**, L25. doi:10.1088/2041-8205/733/2/L25.

Ma, S., Raymond, J.C., Golub, L., Lin, J., Chen, H., Grigis, P., Testa, P., Long, D.: 2011, Astrophys. J. **738**, 160. doi:10.1088/0004-637X/738/2/160.

Magdalenic, J., Vrsnak, B., Pohjolainen, S., Temmer, M., Aurass, H., Lehtinen, N.J.: 2008, Solar Phys. **253**, 305. doi: 10.1007/s11207-008-9220-x.

Magdalenic, J., Marque, C., Zhukov, A. N., Vrsnak, B., Zic, T.: 2010, Astrophys. J. **718**, 266. doi:10.1088/0004-637X/718/1/266.

Mann, G., Classen, T., Aurass, H.: 1995, Astron. Astrophys. **295**, 775.

Mann, G., Klassen, A., Classen, T.: 1996, Astron. Astrophys. **119**, 489.

Nelson, G.J., Melrose, D.B.: 1985, in Solar Radio Physics, edited by D.J. McLean and R.D. Robinson, p 333.

Newkirk, G., Jr. 1961, Astrophys. J. **133**, 983. doi: 10.1086/147104.

Payne-Scott, R., Yabsley, D.E., Bolton, J.G.:1947, Nature, **160**, 256. doi:10.1038/160256b0.

Patsourakos, S., Vourlidas, A., Stenborg, G.: 2010, Astrophys. J. Lett. **724**, L188. doi:10.1088/2041-8205/724/2/L188.

Saito, K.: 1970, Ann. Tokyo Astr. Obs. **12**, 53.

Sittler, E.C., Jr., Guhathakurta, M.: 1999, Astrophys. J. **523**, 812. doi: 10.1086/307742.

Smerd, S.F., Sheridan, K.V., Stewart, R.T.: 1974, In: Newkirk, G.A. (ed.), Coronal Disturbances, IAU Symposium **57**, Reidel, 389.

Smerd, S. F., Sheridan, K.V., Stewart, R.T. 1975, Astrophys. J. Lett. **16**, 23.

Subramanian, K.R., Ebenezer, E., Raveesha, K.H.: 2010, In: S.S. Hasan., R.J. Ruttens (ed.), Magnetic coupling between the Interior and Atmosphere of the Sun, Astrophysics and space science proceedings, Springer Berlin Heidelberg, p.482. doi:10.1007/978-3-642-02859-5_62.

Vršnak, B., Aurass, H., Magdalenic, J., Gopalswamy, N.: 2001, Astron. Astrophys. **377**, 321. doi:10.1051/0004-6361:20011067.

Vršnak, B., Magdalenic, J., Aurass, H., Mann, G.: 2002, Astron. Astrophys. **396**, 673. doi:10.1051/0004-6361:20021413.

Wild, J.P., McCready, L.L.: 1950, Austral. J. Sci. Res. **A3**, 387.